\documentclass[prl,twocolumn]{revtex4}%
\usepackage{amsfonts}
\usepackage{amsmath}
\usepackage{amssymb}
\usepackage{graphicx}%

\begin{document}
\preprint{cond-mat/0602635}
\title{Anisotropic s-wave superconductivity in single crystals CaAlSi from
penetration depth measurements}
\author{Ruslan Prozorov}
\affiliation{Ames Laboratory and Department of Physics and Astronomy, Iowa State
University, Ames, Iowa 50011}
\author{Tyson A. Olheiser}
\affiliation{Department of Physics, University of Illinois at Urbana-Champaign, 1110 W.
Green Street, Urbana, IL 61801}
\author{Russell W. Giannetta}
\affiliation{Department of Physics, University of Illinois at Urbana-Champaign, 1110 W.
Green Street, Urbana, IL 61801}
\author{Kentaro Uozato}
\affiliation{Department of Applied Physics, The University of Tokyo, Hongo, Bunkyo-ku,
Tokyo 113-8656, Japan.}
\author{Tsuyoshi Tamegai}
\affiliation{Department of Applied Physics, The University of Tokyo, Hongo, Bunkyo-ku,
Tokyo 113-8656, Japan.}

\begin{abstract}
In- and out-of-plane London penetration depths were measured in single
crystals CaAlSi ( $T_{c}=6.2$ K and $7.3$ K) using a tunnel-diode resonator. A
full $3D$ BCS analysis of the superfluid density is consistent with a prolate
spheroidal gap, with a weak-coupling BCS value in the $ab$-plane and stronger
coupling along the $c-$axis. The gap anisotropy was found to significantly
decrease for higher $T_{c}$ samples.

\end{abstract}
\date{16 April 2006}
\maketitle

Superconductors with AlB$_{2}$ structure have received increased
attention after the discovery of superconductivity at 39 K and
especially after identification of two distinct gaps in MgB$_{2}$
\cite{muranaka,canfield,bouquet}. It is believed that two gaps
survive, because of reduced interband scattering due to different
dimensionality of 2D $\sigma$ and 3D $\pi$ bands. Investigating
materials with similar crystal and band structure is therefore
important for understanding the mechanism of superconductivity in
this class of hexagonal - layer compounds. In this paper we study
CaAlSi which has been synthesized long ago \cite{bodak}, but in
which superconductivity was discovered only recently \cite{imai2}.
Band structure calculations show highly hybridized three-dimensional
interlayer and $\pi^{\ast}$ bands \cite{mazin,giantomassi}.
Although, most studies of CaAlSi indicate s-wave pairing, deviations
from a single isotropic gap behavior have been reported
\cite{imai2,ghosh,tsuda,kuroiwa,lorenz}. Magnetic measurements
indicate a fully developed s-wave BCS gap \cite{ghosh,imai}.
Angle-resolved photoemission spectroscopy \cite{tsuda} revealed the
same gap magnitude on the two bands with moderate strong coupling
value for the reduced gap, $2\Delta/k_{B}T_{c}=4.2$. Together with
specific heat measurements \cite{lorenz} it provided reliable
evidence for a three-dimensional moderately strong-coupled s-wave
BCS superconductivity. On the other hand, $\mu$SR studies have been
interpreted as evidence of either one highly anisotropic or two
distinct energy gaps \cite{kuroiwa}. Furthermore, 5-fold and 6-fold
stacking sequence of (Al,Si) layers corresponding to two different
values of $T_{c}$ of $\sim 6$ and $\sim 8$ K were found
\cite{sagayama}. Therefore, an experimental study of in- and
out-of-plane superfluid density is needed to understand anisotropic
superconducting gap structure to help understanding the mechanism of
superconductivity in AlB$_{2}$ type compounds.

Single crystals of CaAlSi were grown from Ca:Al:Si (1:1:1) ignots using a
floating zone method as described elsewhere \cite{ghosh3}. Samples have
$T_{c}$ either $6.2$ or $7.3$ K, which is directly related to different
stacking sequences \cite{sagayama}. We measured two of each types of
slab-shaped crystals with typical dimensions $0.3\times0.3\times0.4$ mm$^{3}$.
The penetration depth was measured with an LC tunnel-diode oscillator which is
sensitive to changes in susceptibility of several pico-emu or, equivalently,
to changes in London penetration depth of about $0.3$ \AA \ for our crystals
\cite{prozorov}. The quantitative analysis of the frequency shift depends on
the sample shape and relative orientation of the excitation field, $H_{ac}$,
with respect to the principal axes. Assuming superconducting crystal with
isotropic in-plane response determined by the in-plane penetration depth,
$\lambda_{ab}\left(  T\right)  $ and possible different value of the $c-$axis
penetration depth, $\lambda_{c}\left(  T\right)  $, at least two experimental
arrangements are required to extract $\lambda_{ab}\left(  T\right)  $ and
$\lambda_{c}\left(  T\right)  $ separately. In the $H_{ac}||c-$ axis
orientation, superconducting currents are generated in the $ab-$plane, thus
the susceptibility is determined only by $\lambda_{ab}\left(  T\right)  $ and
the frequency shift, $\Delta f\left(  T\right)  =f\left(  T\right)  -f_{0}$,
is given by
\begin{equation}
\Delta f\left(  T\right)  =\frac{f_{0}V_{s}}{2V_{0}\left(  1-N\right)
}\left[  1-\frac{\lambda\left(  T\right)  }{R}\tanh\left(  \frac{R}%
{\lambda\left(  T\right)  }\right)  \right]  \label{df}%
\end{equation}
where $V_{0}$ is the effective coil volume, $N$ is the demagnetization factor,
$\lambda\left(  T\right)  $ is the London penetration depth, $R$ is the
effective planar sample dimension \cite{prozorov}. With magnetic
susceptibility $\chi$ this equation is just $\Delta f\left(  T\right)
=-4\pi\chi\left(  T\right)  \Delta f_{0}$ where the only sample shape -
dependent parameter, $\Delta f_{0}$, is measured directly by pulling the
sample out of the coil at low temperature.%

\begin{figure}
[ptb]
\begin{center}
\includegraphics[
height=7.3675cm,
width=9.0303cm
]%
{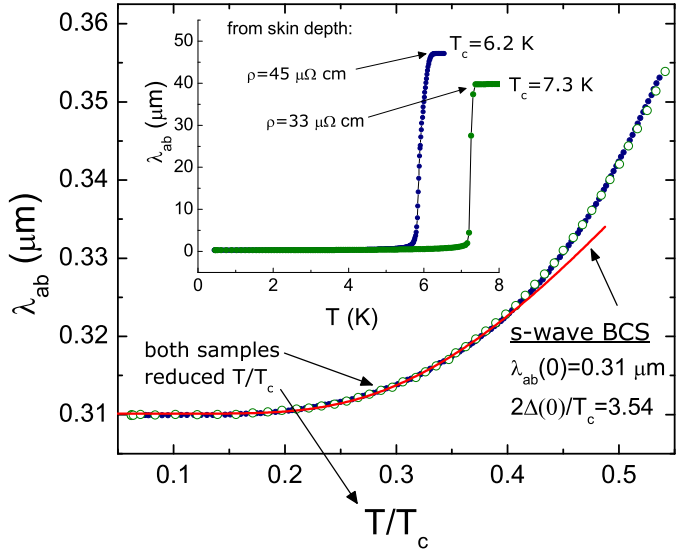}%
\caption{Low-temperature variation of the in-plane penetration depth in two
crystals plotted versus reduced temperature $T/T_{c}$. Red line is a fit to
low-temperature s-wave BCS model, Eq.(\ref{dlBCS}). Inset shows full
temperature scale along with the normal state resistivities extracted from the
skin depth.}%
\label{fig1}%
\end{center}
\end{figure}

Figure \ref{fig1} shows the in-plane penetration depth determined from the
frequency shift using Eq.(\ref{df}). The zero-temperature value was estimated
from the fit to the BCS formula as described below. This value is not
important for the analysis of $\Delta\lambda_{ab}\left(  T\right)  $, but is
needed to estimate the superfluid density. Muon spin rotation gives
$\lambda_{ab}\left(  0\right)  =2390$ \AA \ \cite{kuroiwa}, measurements of
the critical fields $\lambda_{ab}\left(  0\right)  =2060$ \AA , $\lambda
_{c}\left(  0\right)  =870$ \AA \ \cite{imai} and $\lambda_{ab}\left(
0\right)  =3140$ \AA \ \cite{ghosh} as well as measurements of the reversible
magnetization $\lambda_{ab}\left(  0\right)  =3100$ \AA \ \cite{ghosh}. Our
conclusions are not significantly affected by the variation of these values.
The main purpose of Fig.\ref{fig1} is to compare crystals with different
$T_{c}$. The inset shows data on a full temperature scale. The signal
saturates at the level corresponding to the normal-state skin depth, thus
providing additional information - contact-less measurements of the
resistivity above the transition. We find $45$ $\mu\Omega\cdot$cm and $33$
$\mu\Omega\cdot$cm for $6.2$ K and $7.3$ K, respectively, which is in an
agreement with direct measurements on single crystals, which found $36$
$\mu\Omega\cdot$cm on higher $T_{c}$ sample \cite{ghosh2,ghosh3}. On the
contrary, when $\lambda_{ab}\left(  T\right)  $ is plotted versus reduced
temperature $T/T_{c}$, the curves for two samples coincide (no normalization
was done for the $y-$axis). Open symbols in Fig. \ref{fig1} show results for
$T_{c}=7.3$ K crystal, whereas closed symbols show $T_{c}=6.2$ K material. The
data are well fit by the standard weak-coupling s-wave BCS model:%
\begin{equation}
\frac{\Delta\lambda\left(  T\right)  }{\lambda\left(  0\right)  }=\sqrt
{\frac{\pi\Delta\left(  0\right)  }{2T}}\exp\left(  -\frac{\Delta\left(
0\right)  }{T}\right)  \label{dlBCS}%
\end{equation}
where, from $\Delta\lambda_{ab}\left(  T\right)  $, we obtained
$\Delta _{ab}\left(  0\right)  =1.76k_{B}T_{c}$, - a weak-coupling
s-wave BCS superconducting gap.

In the $H_{ac}||ab$ orientation shielding currents flow along both the
$ab-$plane and the $c-$axis. The full magnetic susceptibility is obtained from
the anisotropic London equation, which must be solved numerically to extract
the inter-plane penetration depth $\lambda_{c}\left(  T\right)  $. For a slab
$2b\times2d\times2w$ with magnetic field oriented along the longest side $w$
the following solution has been obtained \cite{mansky}:%
\begin{equation}
\frac{\Delta f_{c}\left(  T\right)  }{\Delta f_{0}^{H||ab}}=1-\frac
{\lambda_{ab}}{d}\tanh\left(  \frac{d}{\lambda_{ab}}\right)  -2\lambda
_{c}b^{2}%
{\displaystyle\sum\limits_{n=0}^{\infty}}
\frac{\tanh\left(  \widetilde{b}_{n}/\lambda_{c}\right)  }{k_{n}^{2}%
\widetilde{b}_{n}^{3}} \label{dfc}%
\end{equation}
\newline where $k_{n}=\pi\left(  n+1/2\right)  $ and $\widetilde{b}_{n}%
=b\sqrt{\left(  \left(  k_{n}\lambda_{ab}/d\right)  ^{2}+1\right)  }.$ Knowing
$\lambda_{ab}\left(  T\right)  $ from independent measurements in the
$H_{ac}||c$ orientation and measuring the total frequency shift upon
extraction of the sample from the coil, $\Delta f_{0}^{H||ab}$, Eq.(\ref{dfc})
is solved numerically to obtain $\lambda_{c}\left(  T\right)  $.%

\begin{figure}
[ptb]
\begin{center}
\includegraphics[
height=7.4092cm,
width=9.1028cm
]%
{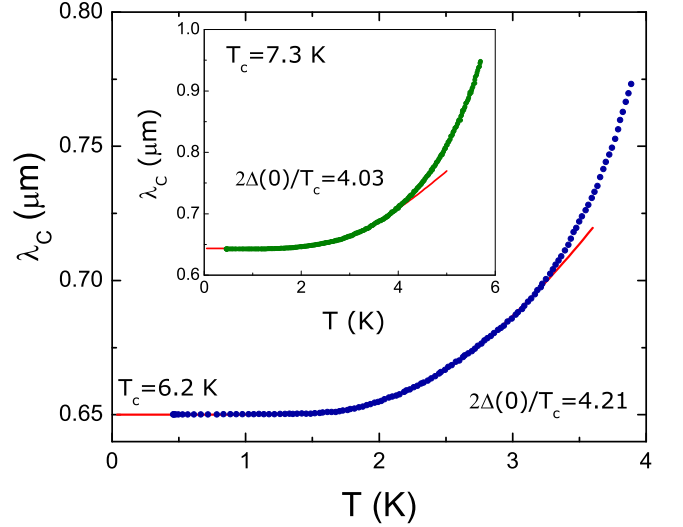}%
\caption{c-axis penetration depth obtained from the numerical inversion of
Eq.(\ref{dfc}) for two different samples. Main frame - for sample with
$T_{c}=6.2$ K and inset shows sample with $T_{c}=7.3$ K. Solid lines are fits
to the low-temperature BCS expression, Eq.(\ref{dlBCS}).}%
\label{fig2}%
\end{center}
\end{figure}

Figure \ref{fig2} shows the out of plane penetration depth obtained
from Eq.(\ref{dfc}). The main frame shows the $T_{c}=6.2$ K sample,
whereas the inset shows data for $T_{c}=7.3$ K sample. Solid lines
are the fits to the low-temperature isotropic BCS expression,
Eq.(\ref{dlBCS}). The gap amplitude obtained from the fits indicates
stronger coupling along the $c-$axis (compared to the
$\lambda_{ab}\left(  T\right)  $ fits, Fig.\ref{fig1}). However, as
shown below, such fitting significantly underestimates the
anisotropy of the superconducting gap. The major problem is that
Eq.(\ref{dlBCS}) is only valid for an isotropic gap that is constant
with temperature, restricting its range of validity to
$T/T_{c}<0.35$. This restriction excludes a significant portion of
the data. A more comprehensive analysis requires a determination of
the normalized superfluid density, which is obtained from the
measured change in the penetration depth, $\Delta \lambda\left(
T\right)  $, via $\rho\left(  T\right)  =\left(  1+\Delta
\lambda\left(  T\right)  /\lambda\left(  0\right)  \right)  ^{-2}.$

The superfluid density generally depends on the shape of the Fermi surface and
the gap anisotropy \cite{Chandrasekhar}. For CaAlSi we can assume a fairly
isotropic Fermi surface \cite{mazin,shein,giantomassi}, a superconducting gap
isotropic in the $ab-$ plane and anisotropic for the out of plane response.
Within the semiclassical approximation \cite{Chandrasekhar},%
\begin{equation}
\rho_{ab}=1-\frac{3}{4T}%
{\displaystyle\int\limits_{0}^{1}}
\left(  1-z^{2}\right)  \left[
{\displaystyle\int\limits_{0}^{\infty}}
\cosh^{-2}\left(  \frac{\sqrt{\varepsilon^{2}+\Delta\left(  z\right)  ^{2}}%
}{2T}\right)  d\varepsilon\right]  dz \label{rhoaa3D}%
\end{equation}

\begin{equation}
\rho_{c}=1-\frac{3}{2T}%
{\displaystyle\int\limits_{0}^{1}}
z^{2}\left[
{\displaystyle\int\limits_{0}^{\infty}}
\cosh^{-2}\left(  \frac{\sqrt{\varepsilon^{2}+\Delta\left(  z\right)  ^{2}}%
}{2T}\right)  d\varepsilon\right]  dz \label{rhocc3D}%
\end{equation}
where $z=\cos\left(  \theta\right)  $ and $\theta$ is the polar angle with
$\theta=0$ along the $c-$axis. Since there is no general argument for the
shape of the gap, we choose the spheroidal form:%
\begin{equation}
\Delta\left(  T,\theta\right)  =\frac{\Delta_{ab}\left(  T\right)  }%
{\sqrt{1-\varepsilon\cos^{2}\left(  \theta\right)  }} \label{gap}%
\end{equation}
and parameter $-\infty\leq\varepsilon\leq1$ is related to eccentricity $e$ as
$\varepsilon=e^{2}=1-c^{-1}$ where $c$ is the normalized semi-axis along the
$c-$axis, but $\varepsilon$ can assume both negative and positive values. The
spheroid is either prolate ($\varepsilon>0$), oblate ($\varepsilon<0$) or a
sphere ($\varepsilon=0$). The temperature dependence of the superconducting
gap was obtained from the anisotropic gap equation. We found that is is well
approximated by $\Delta\left(  T\right)  =\Delta\left(  0\right)  \tanh\left(
1.785\sqrt{T_{c}/T-1}\right)  $.

The following data analysis was performed. By measuring the same sample in two
orthogonal orientations (along the $c-$axis and along the $ab-$plane), both
$\lambda_{ab}\left(  T\right)  $ and $\Delta f_{c}$ were obtained. The latter
contains contributions from both $\lambda_{ab}\left(  T\right)  $ and
$\lambda_{c}\left(  T\right)  $. Equation (\ref{dfc}) was then used to
numerically evaluate $\lambda_{c}\left(  T\right)  $. Low-temperature BCS fits
as well as measurements of the reversible magnetization were used to estimate
$\lambda_{ab}\left(  0\right)  =0.31$ $\mu$m and $\lambda_{c}\left(  0\right)
=0.65$ $\mu$m. Fits over the full temperature range confirmed the assumed
values. Then Eqs. (\ref{rhoaa3D}) and (\ref{rhocc3D}) were used to fit the
data. As a final step both curves, $\rho_{ab}\left(  T\right)  $ and $\rho
_{c}\left(  T\right)  $, were generated from a single set of fitting parameters.%

\begin{figure}
[ptb]
\begin{center}
\includegraphics[
height=7.6245cm,
width=9.1028cm
]%
{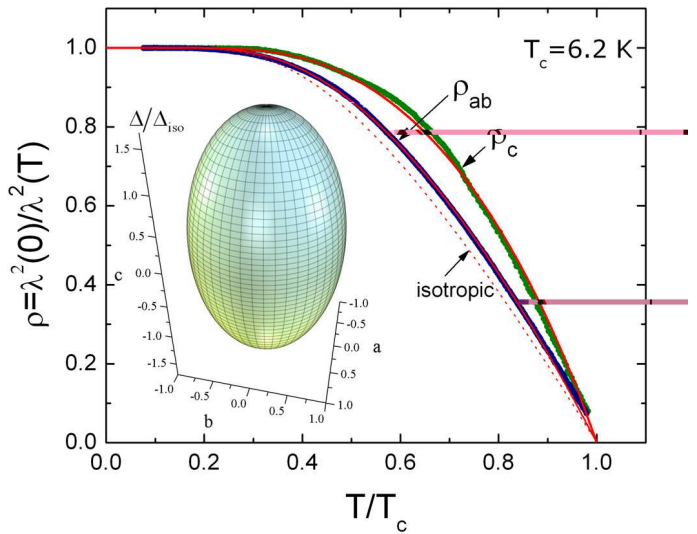}%
\caption{Two components of the superfluid density for samples with $T_{c}=6.2$
K. Symbols show measured data and solid lines corresponding full 3D fiiting.
The dashed line shows isotropic s-wave BCS result. The inset shows ellipsoidal
gap that describes both curves.}%
\label{fig3}%
\end{center}
\end{figure}

Figure \ref{fig3} shows data and fitting results for the lower
$T_{c}=6.2$ K samples. Symbols are the measured data points and
solid lines are calculated for the ellipsoidal gap shown in the
inset. The fitting procedure yielded the weak-coupling BCS value,
$2\Delta_{ab}\left(  0\right)  /k_{B}T_{c}=3.53$, in the $ab-$ plane
and $\varepsilon=0.656$ corresponding to $2\Delta_{c}\left(
0\right)  /k_{B}T_{c}=6.02$ gap maximum along the $c-$axis.%

\begin{figure}
[ptb]
\begin{center}
\includegraphics[
height=7.407cm,
width=8.9029cm
]%
{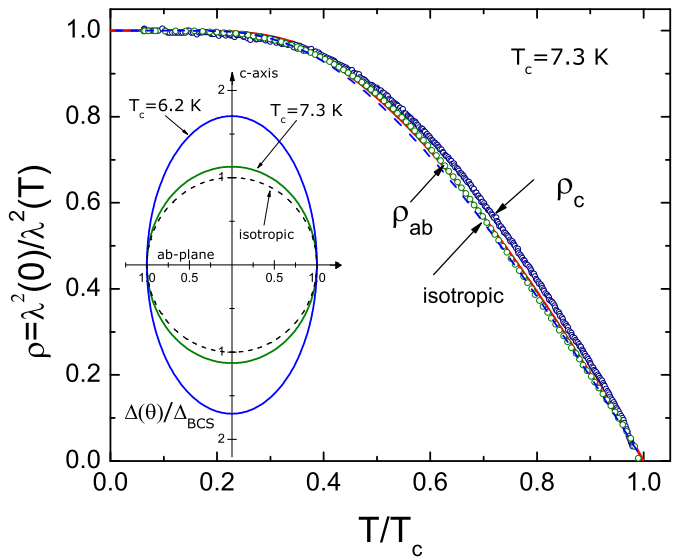}%
\caption{Superfluid densities for the higher $T_{c}=7.3$ K samples
showing apparent reduction of the gap anisotropy copared to Fig.
\ref{fig3}. The
inset compares cross-section of the gap amplitude for the two cases.}%
\label{fig4}%
\end{center}
\end{figure}

Figure \ref{fig4} shows similar results for the samples with
$T_{c}=7.3$ K. There is an obvious reduction of the gap anisotropy.
The best fit to the ellipsoidal gap yields $\varepsilon=0.206$
resulting in $2\Delta_{c}\left( 0\right)  /k_{B}T_{c}=3.98$. It
should be noted that in most previous works only averaged values of
the superconducting gap could be obtained. From heat capacity
measurements, $2\Delta\left(  0\right)  /k_{B}T_{c}=4.07$ was
obtained \cite{lorenz}, whereas ARPES yielded $2\Delta\left(
0\right)  /k_{B}T_{c}=4.2$ \cite{tsuda}. The average effective gap
can be obtained from our results by equating volumes of the
spheroidal gap and a sphere, $\Delta _{eff}=\Delta_{ab}\left(
0\right)  \left(  1-\varepsilon\right)  ^{-1/6}$. This gives
$2\Delta_{eff}\left(  0\right)  /k_{B}T_{c}=4.22$ for samples with
$T_{c}=6.2$ K and $2\Delta_{eff}\left(  0\right)  /k_{B}T_{c}=3.66$
for samples with $T_{c}=7.3$ K which is in the correct range of
reported values and our earlier fits using Eq.(\ref{dlBCS}). All
these values should be compared to the weak-coupling isotropic
result of $2\Delta\left(  0\right) /k_{B}T_{c}=3.53$.

For all samples studied, we find that the temperature dependencies
of both in-plane and out-of-plane superfluid density are fully
consistent with single-gap anisotropic s-wave superconductivity. The
gap magnitude in the $ab-$plane is close to the weak-coupling BCS
value while the $c-$axis values are somewhat larger. Our results
suggest that scattering is not responsible for the difference in
$T_{c}$. Scattering would lead to a suppression of the gap
anisotropy \cite{mazin2}. The gap with average value
$\overline{\Delta}$ and variation  $\delta\Delta$ on the Fermi
surface can only survive if
$\hbar\tau^{-1}\ll\sqrt{\overline{\Delta}\delta\Delta}$, where
$\tau$ is the impurity scattering rate \cite{mazin2} . The values of
resistivity are very close for both high- and low- $T_{c}$ samples
($45$ $\mu\Omega\cdot$cm and $33$ $\mu\Omega\cdot$cm, respectively),
see Fig. \ref{fig1}, the qualitative trend in anisotropy is just
oppisite. Also, 15\% suppression of $T_{c}$ by non-magentic
impurities requires very large concentrations. This would, indeed,
significantly smear the transition, which we did not observe.
Therefore, all facts point out that in CaAlSi gap anisotropy
abruptly decreases as $T_{c}$  abruptly increases from $\sim 6$ to
$\sim 8$ K. A plausible mechanism comes from the analysis of the
stacking sequence of (Al/Si) hexagonal layers \cite{sagayama}. There
are two structures -- 5-fold and 6-fold stacking corresponding to
low and higher - $T_{c}$ samples, respectively. Buckling of (Al,Si)
layers is greatly reduced in a 6-fold structure, which leads to the
enhancement of the density of states, hence higher $T_{c}$. Our
results suggest that reduced buckling also leads to almost isotropic
gap function. This may be due significant changes in the phonon
spectrum and anisotropy of the electron-phonon coupling.

Measurements of the field dependence of the penetration depth in the vortex
state also show a difference between the two sets of samples, as do
measurements taken with different field orientations relative to the $c-$axis.
Of particular interest is the variation of the second critical field and the
so-called peak effect \cite{ghosh,tamegai,tamegai2,ghosh2,ghosh3}.
Tunnel-diode studies of these properties will be reported elsewhere.

We thank Vladimir Kogan and Jules Carbotte for helpful discussions
related to the connection between measured London penetration depth,
anisotropy, strong coupling and two-band superconductivity. Ames
Laboratory is operated for the U.S. Department of Energy by Iowa
State University under Contract No. W-7405-ENG-82. This work was
supported in part by the Director for Energy Research, Office of
Basic Energy Sciences and in part by NSF grant No. DMR-0603841. R.P.
acknowledges support from the Alfred P. Sloan Foundation. RWG and
TAO acknowledge support from the National Science Foundation,
Contract No. DMR 05-03882. KU and TT acknowledge support by
Grant-in-aid from the Ministry of Education, Culture, Sports,
Science, and Technology.

\end{document}